\documentclass[aps,pre,twocolumn,superscriptaddress]{revtex4-2}
\usepackage{graphicx}
\usepackage{amsmath}
\usepackage{mathtools}
\usepackage{siunitx}
\usepackage{tikz}
\usepackage{amsfonts}
\usepackage{CJKutf8}
\usetikzlibrary{shapes,arrows,chains,automata, positioning}
\begin{document}


\title{Superdiffusion in self-reinforcing run-and-tumble model with rests}



\author{Sergei Fedotov}
\email[]{sergei.fedotov@manchester.ac.uk}
\affiliation{Department of Mathematics, University of Manchester, M13 9PL, UK}

\author{Daniel Han}
\email[]{daniel.han@manchester.ac.uk}
\affiliation{Department of Mathematics, University of Manchester, M13 9PL, UK}

\author{Alexey O. Ivanov}
\email[]{ Alexey.Ivanov@urfu.ru}
\affiliation{Ural Mathematical Center, Department of Theoretical and Mathematical Physics, Ural Federal University, Lenin Ave., 51, Ekaterinburg, 620000, Russian Federation}

\author{Marco A. A. da Silva}
\email[]{maasilva@fcfrp.usp.br}
\affiliation{Faculdade de Ci\^encias Farmac\^euticas de Ribeir\~ao Preto, Universidade de S\~ao Paulo (FCFRP-USP), Ribeir\~ao Preto, Brazil}

\date{\today}

\begin{abstract}
This paper introduces a run-and-tumble model with self-reinforcing directionality and rests. We derive a single governing hyperbolic partial differential equation for the probability density of random walk position, from which we obtain the second moment in the long time limit. We find the criteria for the transition between superdiffusion and diffusion caused by the addition of a rest state. The emergence of superdiffusion depends on both the parameter representing the strength of self-reinforcement and the ratio between mean running and resting times. The mean running time must be at least $2/3$ of the mean resting time for superdiffusion to be possible. Monte Carlo simulations validate this theoretical result. This work demonstrates the possibility of extending the telegrapher's (or Cattaneo) equation by adding self-reinforcing directionality so that superdiffusion occurs even when rests are introduced.
\end{abstract}


\maketitle

\section{Introduction}
Persistent random walks with finite velocities are powerful models describing chemotaxis \cite{hillen2002hyperbolic,dolak2003cattaneo,erban2004individual,filbet2005derivation}, organism movement and searching strategies \cite{benichou2011intermittent,mendez2016stochastic,angelani2014first}, intracellular transport \cite{bressloff2013stochastic} and cell motility \cite{selmeczi2005cell,othmer1988models}. 
Stochastic cell movement plays a major role in embryonic morphogenesis, wound healing and tumor
cell proliferation \cite{ridley2003cell}. The modeling 
of cell and bacteria migration towards a favorable environment is usually based on ``velocity-jump'' models describing self-propelled motion with the
runs and tumbles. Finite velocities and inertial resistance to changes in direction make these random walks physically well motivated since random walkers in nature cannot instantaneously jump to different states.
The collective behavior of cells and various organisms is another rapidly growing area of
active matter research \cite{tailleur2008statistical,bechinger2016active}. Various hyperbolic models involving nonlinear partial
differential equations (PDEs) for the population densities have been used for analysis of spatio-temporal patterns describing the chemical and social interactions of organisms \cite{mendez2010reaction,fetecau2010investigation,thompson2011lattice,farrell2012pattern,carrillo2014non}.

Models of cell motility have been predominantly concerned with Markovian random walk models (see for example \cite{othmer1988models,stevens1997aggregation}).
However, the analysis of random movement of metastatic cancer cells shows the anomalous superdiffusive dynamics of cell migration \cite{huda2018levy}. Over the past few years there have been several attempts to model anomalous  transport involving superdiffusion \cite{fedotov2016single,zaburdaev2015levy,estrada2021motility,taylor2016fractional,shaebani2019transient,giona2019age,wang2020large}. Superdiffusion occurs as a result of the  power-law distributed running times with infinite second moment \cite{zaburdaev2015levy} or collective interaction between random walkers \cite{fedotov2017emergence}. Such models are intrinsically non-Markovian involving non-local in time integral terms, making the inclusion of reactions, internal dynamics, chemical signals and inter-particle interactions cumbersome and unwieldy.


Recently, we introduced a persistent random walk model with self-reinforcing directionality that generates superdiffusion from exponentially distributed runs, accurately modelling the statistics found in active intracellular transport \cite{han2021self}. Although this model involves strong memory, it can be formulated as a persistent random walk with space and time dependent coefficients, facilitating convenient implementations of reactions, chemotaxis and interactions using the established methods within the persistent random walk framework.  Self-reinforcing directionality can be introduced using the conditional transition probabilities, $q_{\pm}$, that a particle previously moving in the $+$ and $-$ direction for some random duration chooses to travel in the same direction again for another random duration. Then the simple equations above can be written as
\begin{equation}
	\frac{\partial p_{\pm}}{\partial t}\pm\nu\frac{\partial p_{\pm}}{\partial x} = -\lambda(1-q_{\pm}) p_{\pm}+ \lambda(1-q_{\mp})p_{\mp},
	\label{pplusminus2}
\end{equation} 
where $q_{\pm} =w t^{\pm}/t + (1-w) t^{\mp}/t$. In this case, $t^+$ and $t^-$ are the times that a particle has spent travelling in the positive and negative direction respectively and $t=t^++t^-$. The persistence probability $w$ defines how much the random walk chooses to follow its past behavior. For example, if much time is spent moving in the positive direction ($t_+\rightarrow t$) and $w=1$, then probability that the particle will always choose to move in the positive direction since $q_+\rightarrow1$. The advantage of this formulation is that $q_{\pm}$ can be simply expressed as a function of space, $x$, and time, $t$. If one realizes that $x = \nu(t_+-t_-)$, then 
\begin{equation}
	q_{\pm}(x,t) = \frac{1}{2}\left[1\pm\frac{(2w-1)(x-x_0)}{\nu t}\right]. 
	\label{condtranprob1}
\end{equation}
Expressing $q_{\pm}$ in this way, we can write down \eqref{pplusminus2} as a single hyperbolic PDE, 
\begin{equation}
	\frac{\partial^2 p}{\partial t^2} + \lambda\frac{\partial p}{\partial t} = \nu^2 \frac{\partial^2 p}{\partial x^2} - \frac{ \lambda(2w-1)}{t} \frac{\partial ((x-x_0) p)}{\partial x}.
	\label{masternorests}
\end{equation}
This model has been shown to exhibit superdiffusion despite having exponentially distributed run times \cite{han2021self}. For values of $w>1/2$, the conditional transition probabilities generates self-reinforcing directionality in \eqref{masternorests} and for $w>3/4$, superdiffusion. Research on reinforcement in random walks has been explored in jump processes \cite{stevens1997aggregation}. The model represented in \eqref{masternorests} is actually a continuous space and time generalization of the elephant random walk \cite{schutz2004elephants,da2014ultraslow,boyer2014solvable,baur2016elephant,bercu2019hypergeometric,bercu2019multi,da2020non}, which is discrete in space and time.

A limitation of \eqref{masternorests} is that only active states were included in the model. In reality, most natural phenomena have rest states associated with passive movement, no movement or even death. In particular, animals move by alternating between foraging and resting \cite{mashanova2010evidence,tilles2016random} In modelling processes with rest states, the L\'evy walk with rests \cite{klafter1994levy,zaburdaev2002enhanced,klafter2011first,portillo2011intermittent} and persistent random walks with death \cite{fedotov2015persistent} have been introduced.

The aim of this paper is to formulate the self-reinforcing velocity random walks with stochastic rests. Important questions for this model is: Does superdiffusion still exist after introducing rests? If superdiffusion does exist, what is the critical value of the ratio of mean running and resting time for which the phase transition from diffusion to superdiffusion occurs?

In the first section, we formulate the self-reinforcing directionality random walk with a rest state and derive the non-local hyperbolic governing partial differential equation for the PDF of particle position. In the second section, we find an analytical expression for the second moment and the critical point where the transition from diffusion to superdiffusion occurs. Finally, we present the Monte Carlo simulations of the random walk with reinforcement, which confirms the existence of superdiffusion.

\section{Self-reinforcing directionality with rests}
In this section, we introduce the self-reinforcing random walk with rests that transitions between moving states via an intermediate resting state. Consider a particle that moves with constant velocity $\pm\nu$ in the positive or negative direction for exponentially distributed running times with rate $\lambda$. The resting time is exponentially distributed with rate $\eta$. The governing equations that describe the probability densities $p_+$, $p_-$ and $p_0$ for the rest state are 
\begin{equation}
	\begin{split}
		\frac{\partial p_{\pm}}{\partial t} &\pm \nu \frac{\partial p_{\pm}}{\partial x} = -\lambda p_{\pm} + \eta r_{\pm} p_0, \\
		\frac{\partial p_0}{\partial t} &= \lambda p_+ + \lambda p_- - \eta (1-r_0)p_0,
	\end{split}
\label{masterrests}
\end{equation}
where $r_{\pm}$ are self-reinforcing conditional transition probabilities and $r_++r_-+r_0=1$.

The introduction of self-reinforcing directionality in this paper is done through the conditional transition probabilties,
\begin{equation}
	\begin{split}
		r_{\pm} = w_1\frac{t^{\pm}}{t} + w_2 \frac{t^{\mp}}{t} + w_3 \frac{t^0}{t} \hspace{0.2cm} \text{and} \hspace{0.2cm} r_0 =& w_3,
	\end{split}
\label{condtranprobrests}
\end{equation}
where $r_++r_-+r_0=1$ and $t^+$, $t^-$ and $t^0$ are the relative times that the particle has spent in the positive velocity, negative velocity or resting state, respectively. The weights, $w_1$, $w_2$ and $w_3$, represent the amount of influence that each relative time has on the probability that a particle will transition to the corresponding state. Naturally, the weights are positive and $w_1+w_2+w_3 = 1$. 

Why and how does \eqref{condtranprobrests} introduce self-reinforcing directionality into \eqref{masterrests}? We demonstrate the effect on the conditional transition probabilities by considering weights $w_1$ and $w_2$. For $w_1>w_2$, the random walk reinforces its own past behavior by increasing the transition probability to the positive velocity state, $r_{+}$, when the time spent in that state, $t^{+}$, increases. The same can be said between $r_-$ and $t^-$. In other words, the more the random walk spends time in either the positive or negative velocity state, the more likely a future transition into that state becomes. So the weights $w_1$ and $w_2$ perform an essential function in self-reinforcing directionality by either `punishing' or `rewarding' past choices and making future transitions to states dependent on time spent in the two active states. Now, we present a clear and effective method for simplifying \eqref{condtranprobrests} so that a single governing equation can be obtained.

We can rewrite \eqref{condtranprobrests} using $t=t^++t^-+t^0$ and $x = x_0 +\nu(t^+-t^-)$ as 
\begin{equation}
	\begin{split}
		r_+(x,t) &= \frac{w_1-w_2}{2}\frac{x-x_0}{\nu t}+ \frac{w_1+w_2}{2}+\Lambda\frac{t^0}{t}\\
		r_-(x,t) &=  -\frac{w_1-w_2}{2}\frac{x-x_0}{\nu t}+ \frac{w_1+w_2}{2}+\Lambda\frac{t^0}{t}\\
		r_0& =  w_3, \\
	\end{split}
\label{condtranprobrests2}
\end{equation}
where $\Lambda = -(w_1+w_2)/2+w_3$. In the case where $t_0$ does not contribute to the conditional transition probabilities, $w_3 = 1/3$ and $w_1+w_2 = 2/3$.
Then \eqref{condtranprobrests2} becomes
\begin{equation}
		r_{\pm} = \frac{1}{3}\pm\alpha_0\frac{x-x_0}{2\nu t} \hspace{0.3cm} \text{and} \hspace{0.3cm} r_0 =  \frac{1}{3},
	\label{condtranprobrests3}
\end{equation}
where 
\begin{equation}
	\alpha_0 = w_1-w_2 \text{ and } -2/3<\alpha_0<2/3. 
	\label{alpha_def}
\end{equation}

The formulation of self-reinforcement in this way presents a particularly powerful mechanism to introduce memory effects and superdiffusion. It is clear that this mechanism is different to that used to generate superdiffusion in continuous time random walks or L\'evy walks. Using the definition of conditional transition probabilities in \eqref{condtranprobrests3}, we can formulate a single governing equation that enables various extensions, such as reactions, interactions and chemotaxis, to be readily applied from the persistent random walk framework. In our previous paper, we suggested a simple microscopic mechanism of self-reinforcement (see Section VII `Biological Origins' in \cite{han2021self}).

Now, we will derive the single governing equation. From combining \eqref{masterrests}, we obtain three equations
\begin{equation}
	\begin{split}
		\frac{\partial p}{\partial t}= & -\frac{\partial J}{\partial x} \text{,   }\hspace{0.5cm} \frac{\partial p_0}{\partial t} =  \lambda p - \gamma p_0 \hspace{0.5cm} \text{and}\\
		\frac{\partial J}{\partial t} = & -\nu^2 \frac{\partial p}{\partial x} + \nu^2 \frac{\partial p_0}{\partial x} - \lambda J + \nu\eta \left(r_+-r_-\right)p_0,
	\end{split}
\label{cons_flux_masterrests}
\end{equation}
where $p = p_++p_-+p_0$, $J = \nu p_+ - \nu p_-$ and 
\begin{equation}
	\gamma = \lambda + (1-r_0)\eta.
	\label{gamma_def}
\end{equation}
The initial conditions are
\begin{equation}
	\begin{split}
		p(x,0) = \delta(x-x_0), \hspace{0.5cm} p_0(x,0) = 0\\ 
		\text{and} \hspace{0.5cm} J(x,0) = \nu(2u-1)\delta(x-x_0),
	\end{split}
	\label{initialconds}
\end{equation}
where $u$ is the probability that the particle begins with positive velocity and $(1-u)$ to begin with negative velocity. Solving the second equation in \eqref{cons_flux_masterrests} with the initial condition $p_0(x,0)=0$, one can also write $p_0$ in terms of $p$ as
\begin{equation}
	p_0 = \lambda \int_{0}^{t} e^{-\gamma (t-t')}p(x,t') dt'.
	\label{p0asconvolutionofp}
\end{equation}

Combining \eqref{cons_flux_masterrests}, \eqref{p0asconvolutionofp} and \eqref{condtranprobrests3}, a single equation can be found for $p$ as
\begin{equation}
	\begin{split}
		\frac{\partial^2 p}{\partial t^2} &+ \lambda \frac{\partial p}{\partial t} - \nu^2 \frac{\partial ^2 p }{\partial x^2} + \lambda \nu^2 \int_{0}^{t} e^{-\gamma (t-t')}\frac{\partial ^2 p(x,t')}{\partial x^2} dt'\\
		+&\frac{\lambda \alpha_0\eta}{t}  \frac{\partial}{\partial x}\left[(x-x_0)\int_{0}^{t}e^{-\gamma(t-t')}p(x,t')dt'\right] = 0.
	\end{split}
	\label{mastermasterrests}
\end{equation}
Now the crucial question is: Does the intermediate rest state destroy superdiffusion seen in the self-reinforcing directionality random walk model? To answer this, we perform moment analysis. 

If the parameter $\eta\rightarrow\infty$, then the average rest time, which is $1/\eta$, approaches $0$. The fourth term in \eqref{mastermasterrests} approaches $0$ because $\gamma$ defined in \eqref{gamma_def} $\rightarrow\infty$. However, the last term in \eqref{mastermasterrests} does not approach $0$ because $(1-r_0)\eta e^{-\gamma(t-t')}\rightarrow\delta(t-t')$ as $\eta\rightarrow\infty$. So in this case, \eqref{mastermasterrests} becomes the same as the governing equation in the case of no rests, which can be found in (10) in Ref. \cite{han2021self}.

\section{Moment calculations and Superdiffusion}
To find an analytical expression for the second moment $\mu_2(t) = \int_{-\infty}^{\infty}x^2p(x,t)dx$, we use \eqref{mastermasterrests} with the assumption that $x_0=0$. Then
\begin{equation}
	\begin{split}
		\frac{d^2 \mu_2(t)}{dt^2}& + \lambda\frac{d\mu_2(t)}{dt} - \frac{2\lambda \alpha_0 \eta}{t}\int_{0}^{t}e^{-\gamma(t-t')}\mu_2(t')dt'\\ &= 2\nu^2\left(1-\frac{\lambda}{\gamma}\right) + \frac{2\nu^2\lambda}{\gamma}e^{-\gamma t}.
	\end{split}
\label{secondmomentmaster}
\end{equation}
Now using the initial conditions \eqref{initialconds}, we obtain the initial conditions for the second moment
\begin{equation}
	\mu_2(0) = 0 \hspace{0.2cm} \text{and} \hspace{0.2cm} \frac{d\mu_2(0)}{dt} = 0.
	\label{initialconds_secondmoment}
\end{equation}
Using the Laplace transform of \eqref{secondmomentmaster} and \eqref{initialconds_secondmoment}, the equation for $\hat{\mu}_2(s) = \int_{0}^{\infty} \mu_2(t) e^{-st} dt$ is 
\begin{equation}
	\begin{split}
		\frac{d\hat{\mu}_2}{ds}+& \frac{2s+\lambda+2\lambda\alpha_0\eta(s+\gamma)^{-1}}{s(s+\lambda)}\hat{\mu}_2 =\\ &-\frac{2\nu^2}{s^3(s+\lambda)}\left(1-\frac{\lambda}{\gamma}\right) - \frac{2\nu^2\lambda}{\gamma s(s+\gamma)^{2}(s+\lambda)}.
	\end{split}
	\label{secondmomentmaster_laplace}
\end{equation}

Now let us look at the long time limit ($s\rightarrow0$) for \eqref{secondmomentmaster_laplace}, then
\begin{equation}
    \frac{d\hat{\mu}_2}{ds}+ \frac{1+\frac{2\alpha_0\eta}{\gamma} }{s}\hat{\mu}_2 \approx -\frac{2\nu^2}{s^3\lambda}\left(1-\frac{\lambda}{\gamma}\right) - \frac{2\nu^2}{ s\gamma^{3}}.
	\label{secondmomentmaster_laplaceAsymptoticS}
\end{equation}
When neglecting the rest state, $\eta\rightarrow\infty$ or $\gamma = \lambda + (1-q_0)\eta \rightarrow\infty$, then \eqref{secondmomentmaster_laplaceAsymptoticS} becomes
\begin{equation}
    \frac{d\hat{\mu}_2}{ds}+ \frac{1+2\alpha_0}{s}\hat{\mu}_2 \approx -\frac{2\nu^2}{s^3\lambda}.
	\label{secondmomentmaster_laplaceAsymptoticSbeta}
\end{equation}
The homogeneous solution for \eqref{secondmomentmaster_laplaceAsymptoticSbeta} is $\hat{\mu}_2(s) = C_2s^{-2\alpha_0-1}$ where $C_2$ is a constant, which gives $\mu_2(t) \sim t^{2\alpha_0}$ taking the inverse. This shows that $2\alpha_0$ is the anomalous exponent. Analogously, from \eqref{secondmomentmaster_laplaceAsymptoticS} we obtain
\begin{equation}
    \hat{\mu}_2(s) \sim C_2s^{-\frac{2\alpha_0\eta}{\gamma}-1}
\end{equation}
which gives
\begin{equation}
    \mu_2(t) \sim C_2t^{\frac{2\alpha_0\eta}{\gamma}}.
    \label{msd_asymptotic}
\end{equation}
This shows that even with rests, self-reinforcing directionality is enough to generate superdiffusion. The first momemnt, $\mu_1(t) = \int_{-\infty}^{\infty}xp(x,t)dx$, can be found in a similar way as
\begin{equation}
    \mu_1(t) \sim C_1 t^{\frac{\alpha_0\eta}{\gamma}}
\end{equation}
where $C_1$ is a constant. In the following sections, we confirm superdiffusion through Monte Carlo simulations for both the second moment and the variance $\text{Var}[x(t)] = \mu_2(t) - (\mu_1(t))^2$ (see Figures \ref{fig:msd} and \ref{fig:var}). The Monte Carlo simulation results in Figs. \ref{fig:var} show that $C_1\neq C_2$ (and further that $C_2>C_1^2$) such that the variance is non zero and follows the same time dependence as the second moment. Now, let us consider for what parameter values superdiffusion is achieved.
\begin{figure}
	\centering
	\includegraphics[width=\linewidth]{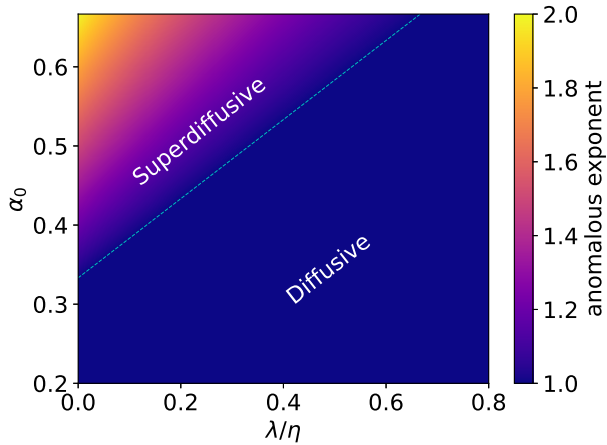}
	\caption{A diagram showing where the diffusive and superdiffusive regimes are found for varying values of $\alpha_0$ and $\lambda/\eta$. The cyan dashed line shows $\alpha_0 = \lambda / 2\eta +1/3$. The anomalous exponent is $2\alpha_0 \eta/\gamma$ in \eqref{msd_asymptotic}.}
	\label{fig:phase_diagram}
\end{figure}

For superdiffusion, the anomalous exponent in \eqref{msd_asymptotic} must satisfy the condition
\begin{equation}
	1< \frac{2\alpha_0}{\frac{\lambda}{\eta}+\frac{2}{3}}<2,
	\label{eq:superdiffusion condition}
\end{equation}
where $\gamma=\lambda + (1-r_0)\eta$ has been used to simplify the expression. Evidently, superdiffusion only depends on two parameters: the self-reinforcement parameter, $\alpha_0$, and the ratio between run and rest rates, $\lambda/\eta$. Rearranging, \eqref{eq:superdiffusion condition} becomes
\begin{equation}
	\frac{1}{3} +\frac{1}{2}\frac{\lambda}{\eta} < \alpha_0 < \frac{2}{3} + \frac{\lambda}{\eta}.
\end{equation} 
The left inequality gives $1/3<\alpha_0$, which in conjunction with \eqref{alpha_def} means that $1/3 < \alpha_0 < 2/3$ is needed for superdiffusion. Then considering \eqref{alpha_def} again, we find the limits $0 \leq \lambda/\eta < 2/3$ are necessary for superdiffusion. The inequality needed for superdiffusion is
\begin{equation}
	\alpha_0 > \frac{1}{2}\frac{\lambda}{\eta} + \frac{1}{3},
\end{equation}
with the bounds $1/3<\alpha_0<2/3$ and $0 \leq \lambda/\eta < 2/3$. The phase diagram showing different parameter values and the corresponding superdiffusive or diffusive states can be seen in Figure \ref{fig:phase_diagram}. 

It is particularly interesting to note that there is a smooth transition from diffusion to superdiffusion dependent on the ratio between running and resting rates, $\lambda/\eta$, in addition to the self-reinforcing parameter, $\alpha_0$. In modelling various different transport phenomena with this self-reinforcing random walk with rests, we expect the dependence of the diffusion-superdiffusion transition on $\lambda/\eta$ to be especially useful as there is a clear physical meaning to why superdiffusion emerges from a random walk with rests. For example, modelling transport mediated by multiple types of motor proteins will involve heterogeneous values of $\lambda$ and $\eta$ and may elucidate why some motor protein transport is more superdiffusive than others.

\section{Monte Carlo simulations}

\begin{figure}
	\centering
	\includegraphics[width = \linewidth]{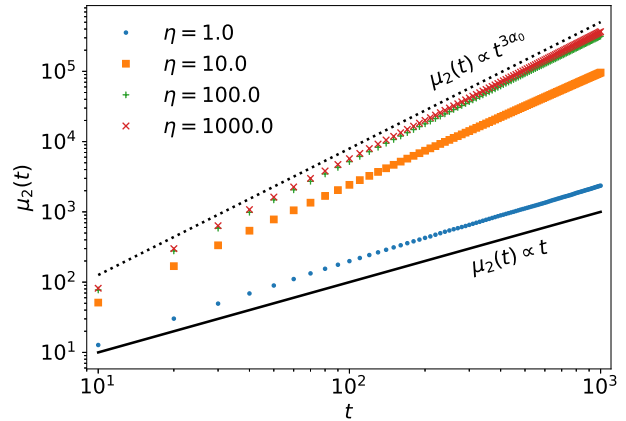}
	\caption{Mean squared displacements for the random walk simulation with varying $\eta$. The parameters for the simulation were $\alpha_0 = 0.6<2/3$, $q_0=1/3$, $\lambda =1$, $\nu=1$ and the number of particles $N=10^4$. The solid black line shows diffusion $\mu_2(t)\sim t$ and dashed black line shows the predicted superdiffusion from \eqref{msd_asymptotic} and \eqref{eq:superdiffusion condition}: $\mu_2(t)\sim t^{3\alpha_0}$ as $\lambda/\eta\rightarrow0$ ($\eta\rightarrow\infty$).}
	\label{fig:msd}
\end{figure}

\begin{figure}
    \centering
    \includegraphics[width=\linewidth]{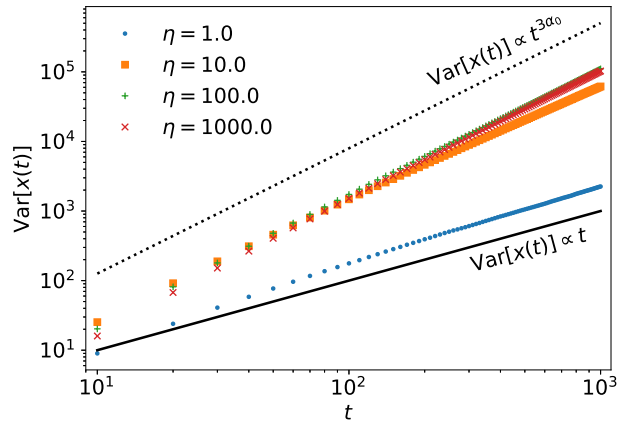}
    \caption{The variance for the random walk simulation with varying $\eta$. The parameters for this simulation were exactly the same as in Fig. \ref{fig:msd}. The solid and dashed black lines are also exactly the same, showing a constant multiplicative difference between the second moment and the variance}
    \label{fig:var}
\end{figure}

\begin{figure}
	\centering
	\includegraphics[width=\linewidth]{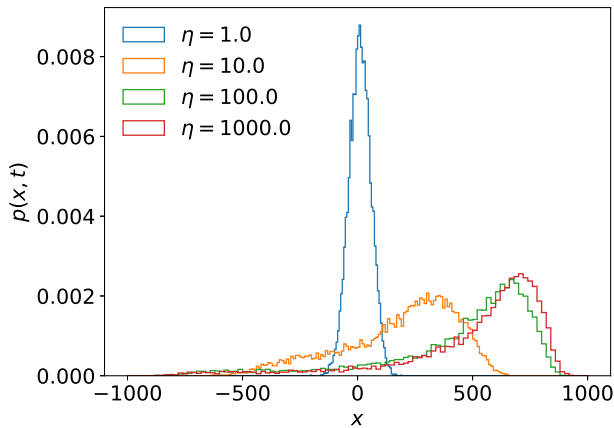}
	\caption{PDF of particle positions at $t=1000$ for the random walk simulation with varying $\eta$. Identical simulation data from Figure. \ref{fig:msd} was used. The parameters for the simulation were $\alpha_0 = 0.6$, $q_0=1/3$, $\lambda =1$, $\nu=1$ and the number of particles $N=10^4$.}
	\label{fig:pdf}
\end{figure}
In this section, we validate the theoretical result in \eqref{msd_asymptotic} and show the displacement PDFs as we vary $\eta$. The numerical simulations for a single random walk corresponding to Eq. \eqref{masterrests} were performed as follows:
\begin{enumerate}
    \item Initialize variables for current simulation time $T_c=0$, particle position $X_c=0$ and current particle state $S_c=1$. In this case, there are only three possible values for $S_c = 0$ or $\pm1$ corresponding to the rest, positive velocity and negative velocity states respectively. For simplicity, we assume the random walk starts in the positive velocity state. 
    \item Initialize the constants of the simulation: $\lambda$, $\eta$, $\nu$, $\alpha_0$, $r_0$ and $t_{end}$, the end time of simulation.
    \item If $S_c = 0$, generate a random number $\Delta T = -\ln(U)/\eta$, where $U\in[0,1)$ is a uniformly distributed random number. If $S_c = \pm1$, generate a random number $\Delta T = -\ln(U)/\lambda$.
    \item Increment the current simulation time $T_c = T_c+\Delta T$ and the particle position $X_c = X_c + \nu S_c \Delta T$. 
    \item If $S_c = \pm1$, set $S_c=0$. If $S_c=0$, generate a uniformly distributed random number, $V\in[0,1)$ and calculate $R_{\pm} = r_0\pm\alpha_0 X_c/(2\nu T_c)$. For $0 \leq V < R_+$, set $S_c = 1$. For $R_+\leq V<R_++R_-$ set $S_c=-1$. Otherwise, set. $S_c=0$.
    \item Iterate steps 3 to 5 until $T_c \geq t_{end}$.
\end{enumerate}
The numerical simulations in this paper was performed using Python3 taking advantage of the `Numba' package for JIT compilation and the `multiprocessing' package for CPU parallelization. These packages were used to significantly improve simulation execution times.

Figures \ref{fig:msd} and \ref{fig:var} show the emergence of superdiffusion and excellent correspondence with \eqref{msd_asymptotic}. Figure \ref{fig:pdf} shows the behavior of the PDF as the value of $\eta$ is varied. Clearly, when the rests become negligible in the asymptotic limit $\lambda/\eta \rightarrow 0$ ($\eta \rightarrow\infty$), the drift of particles caused by self-reinforced directionality dominates. This clearly shows that particles engage in self-reinforcing directionality as rest states become less time consuming and particles choose to move in the same direction as their past history.

\section{Conclusion}

In this paper, we have formulated a run-and-tumble model with self-reinforcing directionality and rests. The system of PDEs \eqref{masterrests} has been reduced to a single, non-local equation for the total probability density \eqref{mastermasterrests}. From this single governing equation, we demonstrated the emergence of superdiffusion by deriving the second moment for the long time limit. This emergence depends on two parameters: the self-reinforcement of particles, $\alpha_0$, and the ratio between running and resting rates, $\lambda/\eta$. We find that at the critical point, $\lambda/\eta = 2/3$, superdiffusion emerges and remains for $\lambda/\eta<2/3$. In other words, the mean running time must be at least $2/3$ of the mean resting time for superdiffusion to occur in this model. Interestingly, we find that even a rest state cannot completely destroy the superdiffusion generated by self-reinforcement. Further, we present the method for Monte Carlo simulation of these random walks and show that the second moment corresponds with theoretical predictions. A natural extension of introducing resting times distributed with constant rate $\eta$ is to introduce a rest state that is non-Markovian with resting times distributed with a residence time dependent rate. This will generate power-law distributed resting times and is a well-known way to introduce subdiffusion in alternating systems \cite{fedotov2010anomalous,fedotov2011non}. It would interesting to find out if this destroys superdiffusion. It is also interesting to consider the case when the velocities alternate at non-exponentially distributed random times \cite{di2001random} or driven by random trials \cite{crimaldi2013generalized}.
Furthermore, this new framework opens new avenues to include interactions of particles by density dependent rates, $\lambda(p)$ and $\eta(p)$, and velocity, $\nu(p)$, leading to aggregation and pattern formation in active matter \cite{bechinger2016active}. 

\begin{acknowledgments}
	S.F is thankful for the support and hospitality of the Ural Mathematical Center at the Ural Federal University, Ekaterinburg. SF also  acknowledges financial support from RSF project 20-61-46013.	D.H. acknowledges the support from Wellcome Trust Grant No. 215189/Z/19/Z. 
	A.O.I. acknowledges financial support from the Ministry of Science and Higher Education of the Russian Federation (Ural Mathematical Center Project No. 075-02-2021-1387). 
	D.H and M.A.A.S acknowledge financial support from FAPESP/SPRINT
	Grant No. 15308-4.
\end{acknowledgments}

\bibliography{real}

\end{document}